\documentclass[12pt]{article}
\usepackage{a4}

\begin{document}

\title{\bf Comment on ``Two-phase behavior of financial markets''}
\author{Marc Potters$^*$ and  Jean-Philippe Bouchaud$^{\dagger,*}$}
\date{\today}
\maketitle

In a recent article \cite{Nature}, the
authors report evidence for an intriguing two-phase behavior of financial
markets when studying the distribution of volume
imbalance ($\Omega$) conditional to the local intensity of its 
fluctuations ($\Sigma$). We show here that this apparent phase transition
is a generic consequence of the conditioning and exists even in the absence
of any non trivial collective phenomenon.

More precisely, to each trade $i$ one can associate the volume of the trade 
$q_i$ and the `sign' of the trade $a_i$, with $a_i=+1$ for buyer initiated
trades, and $a_i=-1$ for seller initiated trades. Then, the volume imbalance 
is defined as:
\begin{displaymath}
\Omega_t = \sum_{i=1}^{N_t} a_i q_i \equiv N_t \langle a_i q_i \rangle,
\end{displaymath}
where $N_t$ is the number of trades taking place in a small time interval
$\Delta t$ around $t$. The intensity of fluctuations, on the other hand, 
is defined as:
\begin{displaymath}\label{sigma}
\Sigma_t = \langle |a_i q_i - \langle a_i q_i \rangle| \rangle.
\end{displaymath}
The finding reported in \cite{Nature} is that conditioned a small value 
of $\Sigma < \Sigma_c$, the probability distribution of $\Omega$ is unimodal, 
and maximum at zero, whereas for $\Sigma > \Sigma_c$, the distribution
becomes bimodal.  
 
Our main point is the following: take a symmetric random variable $X$, 
and an estimate $A$ of its magnitude $|X|$. If we
know that $A$ is small, then $X$ is probably around zero, whereas when
$A$ is known to be large there are two most probable values for 
$X$, namely large and
positive or large and negative. This generic behavior is more precisely 
illustrated on a simple example. If $\Omega$ is an a priori 
Gaussian variable of zero mean and
unit variance and $\Sigma=\beta \Omega^2+\eta$ a noisy estimate of its
magnitude ($\eta$ is a Gaussian noise, independent from $\Omega$, 
of variance $\sigma^2$, and $\beta > 0$)
then:
\begin{displaymath}
P(\Omega|\Sigma)=
\frac{1}{Z(\Sigma)}
\exp\left(-\frac{\Omega^2}{2}-\frac{(\beta \Omega^2-\Sigma)^2}
{2\sigma^2}\right),
\end{displaymath}
in which we recognize the standard `Mexican hat' potential with a
single maximum at zero if $\Sigma<\Sigma_c=\sigma^2/2 \beta$ and two symmetric
maxima $\pm \Omega^*$ otherwise, with $\Omega^* \sim (\Sigma-\Sigma_c)^{1/2}$.
Note that $\Sigma_c$ is finite as soon as $\beta$ is non zero, i.e. if there
are non zero correlations between $\Omega^2$ and $\Sigma$.
This effect is more general and does not require the above normality assumptions.  

Now the local noise intensity $\Sigma$, as defined in Eq. \ref{sigma}
\cite{Nature}, is in fact highly correlated with the magnitude of the volume
imbalance $\Omega$ and therefore the observed `phase transition' is most 
plausibly due to the above mechanism. We have indeed computed the covariance 
between $\Sigma$ and $\Omega^2$
assuming that each trade is independently buyer or seller initiated
with probability one half. We also neglect the fluctuations of $N_t=N$. 
To make the computation more tractable 
without changing 
the conclusion, we define $\Sigma'$ as 
$\langle(q_i a_i-\langle q_i a_i\rangle)^2\rangle$, rather than 
with an absolute value. We find
\begin{displaymath}
\langle\Sigma' \,\Omega^2\rangle_c = 
(N-1)(\langle q^4\rangle-3 \langle q^2\rangle^2)
+ \left(1-\frac{3}{N}\right) \sum_{i \neq j=1}^N 
\langle q^2_i q^2_j \rangle_c ,
\end{displaymath}
where $\langle . \rangle_c$ means connected averages. As soon as 
traded volumes have fat tails 
($\langle q^4\rangle > 3 \langle q^2\rangle^2$) and/or are 
positively correlated ($\langle q^2_i q^2_j
\rangle_c>0$) (see e.g. \cite{Volume}), the last equation shows that
$\Sigma'$ and $\Omega^2$ are indeed positively correlated (i.e. $\beta > 0$). 
Moreover, if
volume correlations are long ranged, the correlation coefficient does not
vanish even for $N$ large.

The above mechanism generically leads to 
$\Omega^*\sim \sqrt{\Sigma-\Sigma_c}$ when
$\log(P(\Omega))$ is smooth at the origin. However choosing
$\log(P(\Omega))\propto-|\Omega|$, a realistic choice for order
imbalance, leads to $\Omega^*\sim {\Sigma-\Sigma_c}$ as observed 
in \cite{Nature}. We therefore suggest that the `two phase' 
behaviour of $P(\Omega|\Sigma)$ is
a direct consequence of known statistical properties of traded volumes.
\vskip 1cm
{\small This comment was submitted to Nature and rejected without refereeing. 
It was deemed too specialized to be of interest for its readership (!). We thank
P. Gopikrishnan, V. Plerou, and H. E. Stanley for a useful correspondence.}
\vskip 1cm

{\small
{$^*$ Science \& Finance, CFM, 109-111 rue Victor Hugo}
{92532 Levallois, France}

{$^\dagger$ Commissariat \`a l'Energie Atomique, Orme des Merisiers}
{91191 Gif-sur-Yvette, France}\\
}


\end{document}